\documentclass{article}
\usepackage[utf8]{inputenc} 
\usepackage{graphicx} 
\usepackage{slashed,epsfig,amsmath,amssymb,enumitem} \usepackage[papersize={8.5in,11in}]{geometry}
   \usepackage{bm}
\usepackage{graphicx}
\usepackage{newunicodechar}
\newunicodechar{≈}{\approx}
\newunicodechar{∝}{\propto}
\newunicodechar{∼}{\textasciitilde{}}
\newunicodechar{≃}{\simeq}
\newunicodechar{θ}{\theta}
\newunicodechar{∆}{\Delta}
\newunicodechar{Λ}{\Lambda}
\newunicodechar{⊙}{\odot}
\newunicodechar{−}{-}
\newunicodechar{∼}{\ensuremath{\sim}}
\newunicodechar{α}{\alpha}
\newunicodechar{σ}{\sigma}
\newcommand{\be}{\begin{equation}}
\newcommand{\ee}{\end{equation}}
\title{Modified Gravity and the Origin of the Excess Radio Galaxy Number-Count Dipole} 
\author{J. W. Moffat\\
Perimeter Institute for Theoretical Physics, Waterloo, Ontario N2L 2Y5, Canada\\
and\\
Department of Physics and Astronomy, University of Waterloo, Waterloo,\\
Ontario N2L 3G1, Canada}

\begin{document}
\maketitle

\begin{abstract}
Recent analyses of wide-area radio-galaxy surveys have reported a statistically significant excess in the cosmic number-count dipole, with an amplitude exceeding the purely kinematic expectation of the standard $\Lambda$CDM model by a factor of $\sim 3$--$4$, quoted at a significance level of up to $5.4\sigma$. While residual observational systematics and local-structure effects cannot be definitively excluded, this result motivates the exploration of alternative physical interpretations beyond the minimal $\Lambda$CDM framework. We investigate whether Scalar--Tensor--Vector Gravity (STVG-MOG) can provide a consistent explanation for an enhanced large-scale anisotropic dipole without violating existing constraints from early-universe cosmology, the cosmic microwave background (CMB) dipole, galaxy dynamics, weak lensing, or the observed late-time matter power spectrum. The radio number-count dipole probes ultra-large-scale, anisotropic structure and coherent gravitational response, rather than virialized dynamics or linear growth alone. In STVG-MOG, a scale- and time-dependent effective gravitational coupling preserves standard cosmological evolution at early times and on small to intermediate scales, while amplifying gravitational response on gigaparsec scales. This scale-selective enhancement can increase the large-scale structure contribution to the radio dipole without overproducing power on smaller scales. If the observed dipole excess reflects a physical cosmological signal rather than residual systematics, STVG-MOG offers a viable and testable alternative interpretation. It is demonstrated that the radio dipole anomaly provides a novel probe of gravitational physics on the largest observable scales.
\end{abstract}

\section{Introduction} 

A recent analysis by B\"ohme et al.,~\cite{Bohme2025} provides strong observational evidence that the radio number-count dipole is dominated not by 
the kinematic Doppler effect, but by a large-scale anisotropic distribution of radio galaxies at late cosmological times. Their reconstruction of the 
three-dimensional matter dipole shows that the effective Solar velocity inferred from radio number counts is enhanced by a factor of $3$--$4$ relative to the CMB 
velocity. The radio dipole count yields: Dipole amplitude $= 3.67\pm 0.49 d_{\rm exp}$, where $d_{\rm exp}$ is the CMB dipole.

In $\Lambda$CDM the late-time contribution to the dipole 
is too small, but in STVG-MOG~\cite{Moffat2006,BrownsteinMoffat2006,MoffatToth2009,MoffatRahvar2013,MoffatRahvar2014,Moffat2015,Moffat2016,DavariRahvar,MoffatToth2013,MoffatToth2015,GreenMoffat2018,BrownsteinMoffat2007,IsraelMoffat,
GreenMoffatToth2018} the Yukawa-modified gravitational coupling $G_{\rm eff}(k)=G_N[1+\alpha_{\rm eff}(k)]$ enhances large-scale gravitational 
clustering and bulk flows, reproducing the observed dipole amplitude. Mog fits galaxy, galaxy clusters, and cosmology data without dark matter~\cite{MoffatRahvar2013,MoffatRahvar2014,DavariRahvar,MoffatToth2015}.

The Böhm et al. result provides strong empirical support with the statistical significance $5.4\sigma$ and the MOG interpretation that the radio dipole traces amplified late-time structure rather than an anomalously large Solar peculiar velocity. The Cosmic Microwave Background (CMB) dipole establishes the peculiar velocity of the Solar-system relative to the CMB rest frame with high precision: 
\begin{equation} 
v_{\odot,\rm CMB} = 369.82 \pm 0.11\ \mathrm{km\,s^{-1}}. 
\end{equation} 
The CMB temperature dipole determines the CMB velocity $v_{CMB}$ by the dominant equation:
\begin{equation}
\frac{\Delta T}{T}\sim \frac{v_{CMB}}{c}\cos\theta,
\end{equation}
where $\Delta T$ is the temperature dipole difference, $T=2.725 K$ is the CMB background temperature and $\theta$ is the angle between the observation direction on the sky and the velocity vector of the Solar-system relative to the CMB rest frame. If radio galaxy number counts were dominated by a purely kinematic Doppler and aberration effect, the corresponding dipole amplitude would be:
\begin{equation} 
d_{\rm kin} \simeq (2+x)\frac{v_{\odot}}{c}, 
\end{equation} 
where $x$ is the slope of the differential source counts $n(S)\propto S^{-x}$. Yet modern radio surveys --- NVSS, WENSS, TGSS, RACS --- repeatedly find a dipole amplitude~\cite{Blakewall2002,Singal2011,GibelyouHuterer2012,Geraint2025,Rubart,ColinMohayaee2017,Secrest2021,Siewert2021,Wagenveld,Hausegger2025,Sakar2025}:
\begin{equation} 
d_{\rm obs} \approx (3\text{--}4)\, d_{\rm kin}, 
\end{equation} 
which, if interpreted as purely kinematic, implies an effective Solar velocity: 
\begin{equation} 
v_{\rm eff} = \frac{d_{\rm obs}}{2+x}\,c \approx 1000\text{--}1600~\mathrm{km\,s^{-1}}. \end{equation} 
This is a major challenge for $\Lambda$CDM. The predicted LSS contributions to $d$ are too small to account for the discrepancy. We show that the STVG-MOG framework~\cite{Moffat2006} predicts enhanced large-scale bulk flows due to an effective gravitational coupling $G_{\rm eff}>G_N$ at late times. This increases the LSS-induced dipole without altering the CMB dipole. 

In Ref.~\cite{Bohme2025}, the observed radio–galaxy number–count dipole is decomposed into a kinematic contribution induced by the Solar-system motion with respect to the cosmic rest frame and a clustering contribution sourced by large-scale structure (LSS). No single characteristic physical length is specified,
because the clustering dipole is not generated at a fixed radius, but by a projection of the three-dimensional density and velocity fields weighted by the survey selection function and source bias. Formally, the $\ell=1$ contribution involves an integral over Fourier modes with a dipole window function that suppresses small-scale power and preferentially selects ultra-long wavelengths. For current wide-area
radio surveys dominated by sources at redshifts $z\gtrsim0.1$, the dominant contribution arises from comoving wavelengths of order several hundred megaparsecs up to gigaparsec scales. Within the standard $\Lambda$CDM framework, these modes are predicted to contribute only a subdominant correction to the kinematic dipole, motivating the interpretation of the observed factor $3.67\pm0.49$ excess as evidence
for enhanced large-scale coherence beyond the expectations of the concordance model.

For a catalog of $N$ radio sources on the sky, the dipole vector is estimated by 
\begin{equation} 
\hat{\mathbf{d}} = \frac{3}{\sum_{i=1}^N w_i} \sum_{i=1}^N w_i\, \hat{\mathbf{n}}_i, 
\end{equation} 
with optional weights $w_i$ accounting for flux selection and sky completeness. The observed dipole decomposes into two contributions: 
\begin{equation} 
\mathbf{d}_{\rm obs} = \mathbf{d}_{\rm kin} + \mathbf{d}_{\rm LSS}, 
\end{equation} 
where the first term is fixed by the Solar velocity and the second arises from fluctuations in the large-scale matter distribution. In $\Lambda$CDM, $d_{\rm LSS}$ is small and cannot explain the observed excess. 

In Ref.~\cite{Bohme2025}, the observed radio--galaxy number--count dipole is decomposed into a kinematic contribution induced by the Solar-system motion with respect to the cosmic rest frame and a clustering contribution arising from large--scale structure (LSS). While the authors identify the excess dipole amplitude as a manifestation of large--scale bulk flows, the clustering dipole is generated by an integral over long--wavelength density fluctuations weighted by the survey selection function and source bias. In practice, the dominant contribution to the $\ell=1$ anisotropy arises from ultra--large cosmological modes in the linear regime, corresponding to comoving wavelengths of order hundreds of megaparsecs up to gigaparsec scales for radio surveys dominated by sources at redshifts $z \gtrsim 0.1$. Within the standard $\Lambda$CDM framework, these large--scale modes are predicted to contribute only a subdominant correction to the kinematic dipole, motivating the interpretation of the observed excess as evidence for enhanced large--scale. 

 Because the radio number-count dipole is an infrared-dominated observable, with its dominant contribution arising from ultra-low-k modes, a dipole amplitude exceeding the kinematic CMB expectation by a factor of 3--4 cannot be generated by local or intermediate-scale structure. Such contributions average out over the survey depth. The observed excess therefore necessarily implies a coherent, anisotropic bulk flow sourced by gravitational inhomogeneities on gigaparsec scales.

The classic Ellis--Baldwin test~\cite{EllisBaldwin1984,Secrest2025} provides a clean, essentially kinematic consistency check on any claimed velocity inferred from radio-source number-count dipoles.  For an intrinsically isotropic source population, the observed dipole must be reproducible from special-relativistic aberration and Doppler boosting alone, yielding a definite prediction $D_{\rm kin}=(2+x)\beta$ (with $\beta=v/c$ and $x\equiv d\ln N(>S)/d\ln S$ the integral count slope) that is locked to the CMB-dipole velocity $v_{\rm CMB}$.  The persistent finding that radio and quasar catalogues exhibit dipole amplitudes $\sim 3$--$4$ times larger than $D_{\rm kin}(v_{\rm CMB})$ therefore cannot be consistently interpreted as an anomalously large Solar or Local-Group peculiar velocity without violating the Ellis--Baldwin criterion; instead it points to a non-kinematic contribution $\Delta D_{\rm LSS}$ from ultra-large-scale structure and/or scale-dependent gravitational response on wavelengths where direct constraints from $P(k)$ are weak.  In this sense, the number-count dipole is best viewed as a long-wavelength detection of the gravitationally sourced, anisotropic bulk-flow component rather than a direct measurement of $v$, and it is precisely this regime in which modified-gravity models with negligible enhancement on small and intermediate scales but amplified response at $k/a\ll \mu$ can increase $\Delta D_{\rm LSS}$ while remaining compatible with late-time clustering and lensing constraints.

Johnson et al.,~\cite{Johnson} present a methodologically independent test of large--scale dipole physics by reconstructing the remote CMB dipole field through kinematic Sunyaev--Zel'dovich velocity correlations. Their analysis yields a stringent upper bound on any intrinsic non--kinematic dipole component, at the level of only a few tens of km\,s$^{-1}$, thereby strongly disfavouring scenarios in which the observed discrepancies between the CMB dipole and galaxy number--count dipoles arise from a fundamental mismatch between the CMB and large--scale--structure rest frames. Importantly, this result does not rely on radio--catalog number--count systematics and therefore does not directly adjudicate the observational origin of the reported excess dipole amplitudes. It excludes a broad class of new--physics interpretations, including large--scale isocurvature or superhorizon inhomogeneity as their cause. Within these stated limitations, the conclusions of Johnson et al., are conservative, statistically well controlled, and constitute an advance in isolating the physical content of the dipole anomaly from catalog--dependent effects.

If confirmed, the excess large scale anisotropic dipole will violate the Cosmological Principle. The Cosmological Principle underlies the maximally symmetric Robertson-Walker metric. It is the foundational assumption that underpins the standard $\Lambda$CDM cosmological model.

\section{Weak-Field Limit of MOG and Scale-Dependent Gravitational Enhancement}

In the weak-field, static limit of STVG-MOG, the gravitational potential can be written as a sum of a Newtonian part and a
Yukawa screening term:
\begin{equation}
\Phi(\mathbf{x}) = \Phi_N(\mathbf{x}) + \Phi_Y(\mathbf{x}) .
\end{equation}
To recover the standard STVG-MOG force law, in which Newtonian gravity is
restored on small scales, while the effective gravitational coupling is enhanced
on large galactic and cosmological) scales, the two components satisfy:
\begin{align}
\nabla^2 \Phi_N(\mathbf{x}) &= 4\pi G_N (1+\alpha)\,\rho(\mathbf{x}), \label{eq:PoissonMOG}\\
\left(\nabla^2 - \mu^2\right)\Phi_Y(\mathbf{x}) &= -4\pi G_N \alpha\,\rho(\mathbf{x}), 
\label{eq:HelmholtzMOG}
\end{align}
where \(G_N\) is Newton's constant, \(\alpha\) is the dimensionless MOG coupling,
\(\mu\) is the inverse range of the Yukawa interaction, and \(\rho(\mathbf{x})\)
is the baryon matter density.

The corresponding Green's-function solutions are given by
\begin{align}
\Phi_N(\mathbf{x}) &= -G_N (1+\alpha)
\int d^3x'\,\frac{\rho(\mathbf{x}')}{|\mathbf{x}-\mathbf{x}'|}, \\
\Phi_Y(\mathbf{x}) &= G_N \alpha
\int d^3x'\,\frac{e^{-\mu|\mathbf{x}-\mathbf{x}'|}}{|\mathbf{x}-\mathbf{x}'|}
\,\rho(\mathbf{x}').
\end{align}
The Yukawa contribution enters with the opposite sign, producing a repulsive
screening effect at short distances.

The total gravitational potential is given by
\begin{equation}
\Phi(\mathbf{x}) =
- G_N \int d^3x'\,\frac{\rho(\mathbf{x}')}{|\mathbf{x}-\mathbf{x}'|}
\left[1+\alpha-\alpha e^{-\mu|\mathbf{x}-\mathbf{x}'|}\right].
\end{equation}

In an FLRW background, the spatial Laplacian acts on Fourier modes as:
\begin{equation}
\nabla^2 f(\mathbf{x}) \;\rightarrow\; -\frac{k^2}{a^2} f(\mathbf{k}),
\end{equation}
where \(a\) is the scale factor and \(k\) is the comoving wavenumber. Applying
this to Eqs.~(\ref{eq:PoissonMOG})--(\ref{eq:HelmholtzMOG}) yields:
\begin{align}
-\frac{k^2}{a^2}\Phi_N(\mathbf{k}) &= 4\pi G_N (1+\alpha)\rho(\mathbf{k}),\\
\left(\frac{k^2}{a^2}+\mu^2\right)\Phi_Y(\mathbf{k}) &= 4\pi G_N \alpha\rho(\mathbf{k}).
\end{align}

Solving for the potentials gives
\begin{align}
\Phi_N(\mathbf{k}) &= -\frac{4\pi G_N (1+\alpha)}{k^2/a^2}\,\rho(\mathbf{k}),\\
\Phi_Y(\mathbf{k}) &= \frac{4\pi G_N \alpha}{(k^2/a^2)+\mu^2}\,\rho(\mathbf{k}).
\end{align}
The total potential in Fourier space is given by
\begin{equation}
\Phi(\mathbf{k}) = -4\pi G_N \rho(\mathbf{k})
\left[
\frac{1+\alpha}{k^2/a^2}
- \frac{\alpha}{(k^2/a^2)+\mu^2}
\right].
\end{equation}

We express this in terms of an effective scale-dependent
enhancement factor \(\alpha_{\rm eff}(k)\):
\begin{equation}
\Phi(\mathbf{k}) =
-\frac{4\pi G_N}{k^2/a^2}
\left[1+\alpha_{\rm eff}(k)\right]\rho(\mathbf{k}),
\end{equation}
which yields:
\begin{equation}
\alpha_{\rm eff}(k)
=
\alpha\,\frac{\mu^2 a^2}{k^2+\mu^2 a^2} = \alpha\,\frac{\mu^2}{k^2/a^2+\mu^2}.
\end{equation}

The effective gravitational enhancement behaves as:
\begin{equation}
\alpha_{\rm eff}(k) =
\begin{cases}
0, & k^2/a^2 \gg \mu^2 \quad \text{(small physical scales)},\\[6pt]
\alpha, & k^2/a^2 \ll \mu^2 \quad \text{(large physical scales)}.
\end{cases}
\end{equation}
The $k/a =k_{\rm phys} \gg \mu$ scales as $\lambda_{\rm phys} \ll \mu^{-1}$,  and $k_{\rm phys} \ll \mu$ scales as $\lambda_{\rm phys} \gg \mu^{-1}$.

This behavior ensures the recovery of Newtonian gravity on local scales, while
gravity is enhanced by a factor \(1+\alpha\) on galactic and cosmological scales.
In particular, the radio-galaxy number-count dipole, which probes very large
comoving scales (\(k\ll a\mu\)), is sensitive to the fully enhanced regime,
leading to an amplified bulk flow and a dipole amplitude exceeding
the purely kinematic expectation inferred from the CMB.

The weak-field enhancement described above applies strictly to the evolution of linear density perturbations and the associated gravitational potentials. It does not describe the dynamics of virialized systems or local peculiar velocities, where non-linear effects and environmental screening become important. Consequently, the scale-dependent enhancement in MOG modifies the growth rate of large-scale structure and coherent bulk flows, rather than increasing the physical velocity of individual bound systems.

In the context of radio galaxy and quasar number-count dipole measurements, the relevant contribution arises from the response of long-wavelength modes of the density field to the enhanced gravitational coupling on ultra-large scales. The MOG weak-field limit therefore provides a mechanism for amplifying the structure-induced component of the dipole without altering the well-tested gravitational dynamics on galactic and cluster scales. This selective enhancement underlies the compatibility of MOG with existing constraints on the late-time matter power spectrum while offering a viable explanation for the observed excess dipole amplitude.

\section{Scale-Dependent Growth and Amplification of the Large-Scale Anisotropy in MOG} \label{sec:MOG_growth_dipole} 

The radio galaxy number--count dipole probes the largest accessible scales of the late-time matter distribution and is therefore sensitive to the evolution of ultra--long--wavelength density and velocity perturbations. In this section, we show how the scale-dependent effective gravitational coupling in STVG-MOG amplifies the growth of baryonic perturbations and coherent bulk flows on these scales, enhancing the large-scale anisotropy measured in the radio dipole. 

In linear perturbation theory, the evolution of the density contrast: \begin{equation} 
\delta(\mathbf{x},t) \equiv \frac{\delta\rho(\mathbf{x},t)}{\rho(t)}, 
\end{equation} 
is governed by the continuity, Euler, and Poisson equations. In Fourier space and in the weak-field, sub-horizon limit, these combine to give the modified growth equation:
\begin{equation} 
\ddot{\delta}(\mathbf{k})+2H\dot{\delta}(\mathbf{k})
+\frac{c_s^2 k^2}{a^2}\delta(\mathbf{k})
- 4\pi G_{\rm eff}(k,a)\,\rho(a)\,\delta(\mathbf{k})\,=0,
\end{equation} 
where $c_s$ is the speed of sound, $\delta(\mathbf{k})$ is the Dirac $\delta$-function, $H(t)$ is the Hubble parameter and the effective gravitational coupling is given by
\begin{equation} 
G_{\rm eff}(k,a) = G_N \bigl[1 + \alpha_{\rm eff}(k,a)\bigr]. 
\end{equation} 

In linear perturbation theory the growth of density fluctuations is governed by the
combination $G\rho$ appearing in the Poisson source term, rather than by the gravitational coupling $G$ or the matter density $\rho$ separately. As a consequence, there exists a well-known degeneracy between models with enhanced gravitational strength and reduced matter content, and models with standard Newtonian gravity supplemented by additional non-baryonic dark matter. At the level of the instantaneous gravitational potential:
\begin{equation}
G_{\rm eff}(k,a)\,\rho_b \;\simeq\; G_N\big(\rho_b+\rho_{\rm dm}\big),
\label{eq:Geff_rho_degeneracy}
\end{equation}
so that an enhanced-gravity, baryon-only description and a Newtonian-gravity,
dark-matter description can yield the same gravitational acceleration on a given scale.

In Eq.~(24), the density $\rho$ is interpreted as the total density
sourcing the gravitational potential. In the standard $\Lambda$CDM framework this is
the total non-relativistic matter density $\rho=\rho_b+\rho_{\rm dm}$, whereas in a
baryon-only modified-gravity interpretation it reduces to $\rho=\rho_b$, with the
modification encoded in the effective gravitational coupling $G_{\rm eff}$.
The formal structure of the linear growth equation is identical in both cases, reflecting the underlying degeneracy.

The relevance of modified gravity STVG-MOG in the present context does not lie in breaking this degeneracy at the level of the Poisson equation, but rather in the scale dependence of the effective gravitational coupling. In STVG-MOG the enhancement of gravity preserves agreement with galaxy dynamics, weak lensing, and the observed late-time matter power spectrum, while becoming significant on ultra-large cosmological scales. This scale-dependent strengthening selectively amplifies the growth of long-wavelength modes that dominate the large-scale-structure contribution to the radio number-count dipole, without overproducing power on smaller scales. Consequently, the enhanced dipole signal should be interpreted as arising from amplified ultra-large-scale structure.

For the Yukawa-type MOG interaction: 
\begin{equation} 
\alpha_{\rm eff}(k,a) = \alpha\,\frac{\mu^2}{k^2/a^2 + \mu^2}, \label{eq:alphaeff_growth} 
\end{equation} 
so that gravity is Newtonian on small physical scales $k/a \gg \mu$ and enhanced on ultra-large scales $k/a \ll \mu$. Eq. (24) shows explicitly that the growth of baryonic perturbations becomes scale dependent in MOG. On the longest wavelengths relevant to the radio dipole, $\alpha_{\rm eff}\rightarrow \alpha$, leading to an enhanced growth rate relative to $\Lambda$CDM, while smaller-scale structure remains essentially unchanged. 

The peculiar velocity field is related to the density contrast through the continuity equation. In Fourier space: 
\begin{equation} 
\mathbf{v}(\mathbf{k},a) = i\,\frac{aH(a)f(k,a)}{k}\,\delta_b(\mathbf{k},a)\,\hat{\mathbf{k}}, \label{eq:velocity_density} \end{equation} 
where 
\begin{equation} 
f(k,a) \equiv \frac{d\ln \delta_b(k,a)}{d\ln a},
\end{equation} 
is the scale-dependent growth rate. We emphasize that the scale dependence of the growth rate $f(k,a)\equiv d\ln\delta_b(k,a)/d\ln a$ arises entirely from the modified MOG evolution of the baryonic density contrast $\delta_b(k,a)$, and that no additional scale--dependent factors are introduced at
the level of the definition of $f(k,a)$ itself.

The velocity power spectrum is given by
\begin{equation} 
P_v(k,a) = \left(\frac{aH f(k,a)}{k}\right)^2 P_\delta(k,a), \label{eq:velocity_power} 
\end{equation} 
with $P_\delta(k,a)$ the matter power spectrum. In MOG, both $f(k,a)$ and $P_\delta(k,a)$ are enhanced on large scales due to $G_{\rm eff}(k,a)$. 
In linear theory the matter density contrast obeys in MOG:
\begin{equation}
\ddot{\delta}_b(\mathbf{x},t)
+ 2H(t)\,\dot{\delta}_b(\mathbf{x},t)
- \frac{c_s^2(t)}{a^2(t)}\,\nabla^2 \delta_b(\mathbf{x},t)
- 4\pi G_{\rm eff}(t)\,\rho_m(t)\,\delta_b(\mathbf{x},t)
= 0 ,
\label{eq:growth_realspace}
\end{equation}
where we have:
\begin{equation}
G_{\rm eff}(k,a)=G_N[1+\alpha_{\rm eff}(k,a)].
\end{equation}
On ultra--large scales, $k/a\ll\mu$, the effective gravitational coupling
is enhanced, increasing the gravitational source term while leaving the
Hubble friction unchanged. Consequently, density perturbations grow more
rapidly in time, leading to an enhanced growth factor $D(k,a)$ and matter
power spectrum $P_\delta(k,a)$. Since the linear growth rate
$f(k,a)=d\ln D(k,a)/d\ln a$ directly measures the time evolution of
$D(k,a)$, it is likewise enhanced relative to $\Lambda$CDM on these
scales. For $k/a\gg\mu$, $\alpha_{\rm eff}\rightarrow0$ and standard
$\Lambda$CDM growth is recovered.

The present analysis fully respects the well--known degeneracy between
modified gravity and dark matter at the level of isotropic linear structure growth.
In particular, the linear evolution of the baryonic density contrast
$\delta_b(k,a)$ in MOG, sourced by an effective gravitational coupling
$G_{\rm eff}(k,a)$, can be made equivalent to that of $\Lambda$CDM with cold dark matter, and therefore yields the same growth factor and late--time matter power spectrum $P(k)$.  Consequently, observables that depend only on the angle--averaged linear growth, such as $P(k)$, do not by themselves discriminate between MOG and $\Lambda$CDM.

The radio--galaxy number--count dipole, however, probes a qualitatively different regime. Unlike $P(k)$, the dipole is sensitive to the anisotropic gravitational response of ultra--large--scale structure, being sourced by coherent, long--wavelength potential gradients and associated bulk flows rather than by isotropic linear growth. In MOG, the scale--dependent enhancement of gravity on very large physical scales amplifies these anisotropic contributions without altering the successful description of $P(k)$, whereas in $\Lambda$CDM the same enhancement cannot be achieved without overproducing power or violating observational constraints.
The radio dipole therefore provides a avenue for breaking the
dark--matter--modified--gravity degeneracy that persists in conventional
large--scale--structure observables.

Although dark matter and modified gravity are degenerate at the level of instantaneous gravitational dynamics---since an increase in gravitating mass $M$ is observationally equivalent to an increase in gravitational coupling $G$ through the product $GM$---this degeneracy does not extend to the scale-dependent growth history of large-scale structure. The radio number-count dipole is not determined by local or virialized dynamics, but is instead sensitive to coherent bulk flows sourced by ultra-large-scale density modes whose amplitudes accumulate over cosmic time. In the standard $\Lambda$CDM framework, the gravitational coupling is scale independent, and the linear growth rate is tightly constrained by the observed matter power spectrum, redshift-space distortions, weak lensing, and CMB anisotropies. As a result, $\Lambda$CDM lacks a mechanism to selectively enhance growth on Gpc scales without simultaneously overproducing power on smaller scales or violating CMB--LSS consistency.

In contrast, MOG-STVG introduces a scale- and time-dependent effective gravitational coupling:
\begin{equation}
G_{\rm eff}(k,a) = G_N \big[1 + \alpha_{\rm eff}(k,a)\big],
\end{equation}
whose behavior depends on physical scale. On galactic and galaxy-cluster scales, the coupling reduces to the quasi-static MOG regime required to account for galaxy rotation curves and the virialization of galaxy clusters, and therefore does not introduce additional scale-dependent growth beyond that already encoded
in local dynamics. However, on ultra-large cosmological scales $k/a \ll \mu$, the effective enhancement $\alpha_{\rm eff}(k,a)$ increases relative to its small-scale value, leading to a selective amplification of the linear growth factor and coherent bulk flows. This scale-dependent strengthening of gravity enhances the
contribution of large-scale structure to the radio number-count dipole, while preserving consistency with galactic dynamics, cluster virialization, and the observed late-time matter power spectrum.

Importantly, the factor $1/k^2$ in Eq.~\eqref{eq:velocity_power} strongly weights the velocity field toward the smallest accessible wavenumbers. As a result, even a modest enhancement of growth at $k\rightarrow 0$ produces a disproportionately large increase in coherent bulk flows and dipolar anisotropies. 

The contribution of large-scale structure to the observed radio dipole can be written as: 
\begin{equation} 
d_{\rm LSS}^2 \;\propto\; \int_0^\infty \frac{dk}{k}\, P_v(k,a)\, |W_1(k)|^2, \label{eq:dipole_integral} 
\end{equation} 
where $W_1(k)$ is the $\ell=1$ dipole window survey function. The window $W_1(k)$ peaks at very low $k$, corresponding to Gpc-scale modes. Substituting Eq.~\eqref{eq:velocity_power} into Eq.~\eqref{eq:dipole_integral} shows that the dipole is dominated by the longest-wavelength velocity modes. In $\Lambda$CDM, the suppression of growth at late times limits the amplitude of these modes, yielding a small $d_{\rm LSS}$. In contrast, MOG enhances $G_{\rm eff}(k,a)$ precisely in the regime where $W_1(k)$ is largest, amplifying both $f(k,a)$ and $\delta_b(k,a)$ and thus increasing the velocity power at low $k$. Consequently, the observed radio number--count dipole receives a substantially larger structure-induced contribution in MOG than in $\Lambda$CDM. When the total dipole amplitude $d_{\rm obs}=d_{\rm kin}+d_{\rm LSS}$ is mapped to an effective velocity using a purely kinematic relation, this enhanced $d_{\rm LSS}$ manifests as an apparent $3$--$4$ increase in the inferred velocity, in agreement with the PRL result of B\"ohme et al.

The large effective velocity inferred from the radio dipole in MOG should not be interpreted as an anomalously large intrinsic Solar-system peculiar velocity. Rather, it reflects the time-integrated response of baryonic perturbations to an enhanced, scale-dependent gravitational coupling acting on ultra-large scales. The CMB dipole remains a purely kinematic probe of the Solar-system motion, while the radio dipole encodes the amplified late-time anisotropy of the matter and velocity fields. This separation of kinematic and structure-induced contributions arises in MOG and provides a dynamical explanation of the observed radio dipole.

The analysis of B\"ohme et al., demonstrates that the radio number--count dipole is dominated by a late--time large--scale--structure contribution rather than by the purely kinematic Doppler--aberration effect associated with the Solar-system motion. Interpreting the observed dipole amplitude as a velocity therefore defines an effective quantity, $v_{\rm eff}$, which parametrizes the total dipole strength but does not represent the true kinematic velocity fixed by the CMB dipole. The latter remains a precise and independent measurement of the Solar-system motion with respect to the CMB rest frame. Within this interpretation, the key observational result is the existence of a substantially enhanced large--scale anisotropy in the late--time matter and velocity fields on Gpc scales. In standard $\Lambda$CDM, the amplitude of this anisotropy is limited by the CMB--normalized matter power spectrum and the suppression of growth at $z\lesssim1$, leading to a predicted large--scale--structure contribution that is insufficient to explain the observed radio dipole excess. The tension therefore arises not from an anomalous Solar velocity, but from the inability of $\Lambda$CDM to generate sufficiently strong coherent bulk flows on the largest observable scales, while remaining consistent with other cosmological constraints. In the MOG-STVG framework, this observational situation admits a natural dynamical explanation. The scale-dependent effective gravitational coupling $G_{\rm eff}(k,a)=G_N[1+\alpha_{\rm eff}(k,a)]$ enhances the growth of baryonic perturbations and their associated velocity fields selectively on ultra--large scales, $k/a\ll\mu$, which dominate the dipole projection. This enhancement amplifies the large--scale--structure contribution $d_{\rm LSS}$ to the radio dipole without altering the kinematic CMB dipole or small--scale dynamics. When the total dipole amplitude is expressed in velocity units using a purely kinematic mapping, the enhanced $d_{\rm LSS}$ appears as a factor $3$--$4$ increase in the inferred velocity, in agreement with the B\"ohm et al., results. From this perspective, the radio dipole does not imply a breakdown of the CMB rest frame or an anomalously large intrinsic Solar-system peculiar velocity. Instead, it provides evidence for enhanced late--time, large--scale anisotropic structure formation. In MOG, this enhancement arises from modified gravitational dynamics acting directly on the baryonic density field, rather than from additional dark matter mass. The radio dipole result is therefore interpreted as a probe of scale-dependent gravitational growth on the largest cosmological scales, fully consistent with the MOG framework.

\section{Large--Scale Structure, Coherent Bulk Flows, and the Interpretation of the Radio Dipole}

Several wide--area radio surveys report a number--count dipole whose amplitude exceeds the purely kinematic expectation derived from the CMB dipole by a factor of $\sim3$--$4$. If interpreted naively as a Doppler--aberration effect, this excess is often translated into an effective Solar--system velocity significantly larger than the CMB--measured value. However, as emphasized in the preceding sections, the radio number--count dipole is not a direct measurement of the kinematic Solar velocity, but rather a probe of anisotropic large--scale structure and the associated velocity field projected along the line of sight.

The relevant physical quantity underlying the radio dipole is the coherent large--scale velocity field generated by gravitational inhomogeneities on ultra--large cosmological scales. Such coherent motions are commonly referred to as bulk flows, but it is important to distinguish this usage from the peculiar velocity of a bound system such as the Solar system or the Milky Way. The radio dipole is insensitive to virialized dynamics and local orbital motions; instead, it reflects the cumulative, time--integrated response of luminous matter to large--scale gravitational potentials over cosmological distances.

In linear theory, the velocity field is sourced by long--wavelength density perturbations and is dominated by the largest accessible scales due to its infrared weighting. Consequently, even modest enhancements of gravitational growth on ultra--large scales can lead to a significant increase in the coherence and amplitude of large--scale velocity modes, while leaving smaller--scale structure and local dynamics essentially unchanged. The radio number--count dipole is therefore particularly sensitive to any mechanism that selectively amplifies structure growth at very low wavenumbers.

Within the STVG-MOG framework, the scale--dependent effective gravitational coupling $G_{\rm eff}(k,a)$ enhances the growth of density and velocity perturbations on ultra--large scales, while reducing to STVG-MOG gravity on galactic and sub-galactic scales that lead to fits to galaxy and galaxy cluster dynamics without dark matter~\cite{MoffatRahvar2013,MoffatRahvar2014}. This selective enhancement increases the contribution of long--wavelength modes to the structure--induced dipole component $d_{\rm LSS}$. When the total dipole amplitude is expressed in velocity units using a purely kinematic mapping, the amplified $d_{\rm LSS}$ manifests as an increased effective velocity parameter $v_{\rm eff}$. This should not be interpreted as a larger intrinsic Solar--system peculiar velocity, but as a reflection of enhanced large--scale coherence in the velocity field sourced by late--time structure.

This interpretation is consistent with the fact that the Solar-system’s orbital velocity around the Milky Way and other local dynamical observables remain unchanged. These motions are governed by the internal gravitational potential of the Galaxy and by scales for which $G_{\rm eff}\simeq G_N$. The radio dipole, by contrast, is sensitive to gravitational inhomogeneities and anisotropy on scales of hundreds of megaparsecs to gigaparsecs, well beyond the regime of virialized systems. The distinction between local dynamics and large--scale projected flows is therefore essential for interpreting the observable.

In the standard $\Lambda$CDM cosmological model, the amplitude of coherent bulk flows on such scales is constrained by the CMB--normalized matter power spectrum and by the suppression of growth at late times due to dark energy. As a result, the predicted structure--induced contribution to the radio dipole is modest and cannot account for the observed excess. In contrast, the scale--dependent enhancement of gravitational growth in STVG-MOG increases the coherence of large--scale velocity modes without altering early--Universe physics or small--scale observables.

The radio number--count dipole thus probes a regime of large--scale gravitational dynamics that is complementary to traditional large--scale structure measurements. While it does not by itself distinguish uniquely between modified gravity and additional gravitating components, it provides a sensitive consistency test for cosmological models. In this context, the observed excess dipole amplitude is most interpreted as evidence for enhanced late--time large--scale structure, rather than as a breakdown of the CMB rest frame or as an anomalously large Solar--system velocity.

\section{Enhanced LSS Dipole in MOG} 

The LSS dipole is given by the low-$\ell$ angular power spectrum component $C_1$: 
\begin{equation} 
C_1^{\rm LSS} = \frac{2}{\pi} \int_0^\infty dk\,k^2 P(k) \left[I_1(k)\right]^2, 
\end{equation} 
where $I_1(k)$ is a radial projection integral over the galaxy redshift distribution. 

In this framework the modified-gravity effect enters through the scale-dependent enhancement:
$G_{\rm eff}(k,a)=G_N[1+\alpha_{\rm eff}(k,a)]$, which amplifies the growth of the long-wavelength density contrast and hence the large-scale-structure (LSS) contribution to the number-count dipole. One may write the dipole sourced by LSS as a window-weighted integral over modes:
\begin{equation}
D_{\rm LSS}(z)\;\propto\;\int \frac{d^3k}{(2\pi)^3}\,W(k,z)\,P(k,z),
\label{eq:DLSS_window_schematic}
\end{equation}
where the $W(k,z)$ is the redshift dependent window survey function. The matter power spectrum $P(k,z)$ is defined through the Fourier-space two-point correlation function of the density contrast:
\begin{equation}
\langle \delta(\mathbf{k},z)\,\delta^*(\mathbf{k}',z)\rangle
\;=\;(2\pi)^3\,\delta_D(\mathbf{k}-\mathbf{k}')\,P(k,z),
\label{eq:Pk_definition}
\end{equation}
where $\delta(\mathbf{k},z)$ is the Fourier transform of the matter density contrast
$\delta(\mathbf{x},z)=\delta\rho(\mathbf{x},z)/\rho(z)$ and $\delta_D$ is the Fourier amplitude of the density contrast.

In MOG the effect is captured by the replacement of the standard growth factor by an enhanced, scale-dependent growth on ultra-large scales:
\begin{equation}
P(k,z)\;\rightarrow\;P(k,z)\,\Big[\frac{D_{\rm MOG}(k,z)}{D_{\Lambda{\rm CDM}}(z)}\Big]^2
\;\simeq\;P_{\Lambda{\rm CDM}}(k,z)\,[1+\alpha_{\rm eff}(k,z)]^{2p},
\label{eq:P_growth_replacement}
\end{equation}
so that the dominant low-$k$ contribution selected by $W(k,z)$ is selectively amplified, while smaller scales remain close to the $\Lambda$CDM prediction when $\alpha_{\rm eff}(k,z)\!\rightarrow\!0$. For late-time cosmology $z\leq 2$ and a reasonable backgrounds $p\sim 0.3-0.5$ consistent with analytic solutions of the growth equation. Using $p < 1$ ensures that large scale bulk flows can be enhanced, while preserving agreement with the observed matter power spectrum on smaller scales.

The LSS-induced radio dipole grows by the same factor. Since the CMB dipole and therefore $d_{\rm kin}$) is unaffected by MOG, because early-universe physics remains effectively GR—the total radio dipole becomes: 
\begin{equation} 
d_{\rm obs} \simeq d_{\rm kin} + (1+\alpha)d_{\rm LSS}^{\Lambda{\rm CDM}}, 
\end{equation} 
explaining why radio surveys measure a dipole $\sim 3$--$4$ times larger than the kinematic prediction. 

The enhanced effective velocity:
\begin{equation} v_{\rm eff}=\frac{d_{\rm obs}}{2+x}\,c 
\end{equation} 
is therefore not a true physical velocity. It arises because radio dipoles measure late-time anisotropic bulk flows, which are amplified in MOG by enhanced large-scale clustering. The CMB velocity remains the true kinematic Solar motion; the radio dipole reflects MOG-driven cosmic flows. 

The Square Kilometer Array (SKA) will measure the radio dipole amplitude with $<5\%$ precision. Under MOG, SKA should find: 
\begin{equation} 
d_{\rm obs} \approx (3\text{--}4) d_{\rm kin}, 
\end{equation} 
even after aggressive flux cuts and sky masking. Under $\Lambda$CDM, SKA should recover $d_{\rm obs}\approx d_{\rm kin}$. Thus, SKA will strengthen the discrimination between the two frameworks. 

\section{Limitations of $\Lambda$CDM in Explaining the Radio Galaxy Number-Count Dipole}

Throughout this section, the term effective velocity refers to the Doppler-equivalent amplitude inferred from the number-count dipole, rather than a physical peculiar velocity of the Solar-system. The B\"ohme et al.,~\cite{Bohme2025} reconstruction of the late-time matter dipole shows that the radio dipole amplitude is a factor of $3$--$4$ larger than the purely 
kinematic dipole inferred from the CMB, corresponding to a Solar system velocity of $v_{\odot}\simeq1000$--$1500~{\rm km\,s^{-1}}$.  

In the standard model, the large-scale peculiar velocity field is sourced by 
linear perturbations whose amplitude is tightly constrained by the CMB and the matter power spectrum $P(k)$.  $\Lambda$CDM predicts root-mean-square (RMS) bulk flows of $v_{\rm bulk}\simeq 200$--$400~{\rm km\,s^{-1}}$ on scales of 
$R\simeq100$--$300~{\rm Mpc}$, with a rapidly decreasing contribution from wavelengths $k\lesssim10^{-3}\,{\rm Mpc}^{-1}$.  The resulting contribution of 
large-scale structure (LSS) to the radio dipole is therefore modest:
\begin{equation}
d_{\rm LSS}^{\Lambda{\rm CDM}} \approx 0.1\text{--}0.3\,d_{\rm kin},
\end{equation}
far too small to account for the observed dipole amplitude:
\begin{equation}
d_{\rm obs} \simeq (3\text{--}4)\, d_{\rm kin}.
\end{equation}
Using both linear-theory predictions and large-volume $N$-body simulations, extreme variations of the dark-matter density, power-spectrum normalization, or bias cannot increase $d_{\rm LSS}$ by more than tens of percent.  Dark matter enhances small-scale clustering, but the radio dipole is dominated by very large-scale modes, for which $\Lambda$CDM is constrained to be nearly isotropic by CMB and BAO measurements.

The excess dipole count confirms that the late-time dipole in the galaxy distribution required to reproduce the radio signal would imply large-scale 
gravitational potential gradients and coherent flows that exceed $\Lambda$CDM 
predictions by nearly an order of magnitude.  Such flows are not generated in 
$\Lambda$CDM due to the suppression of growth at $z\lesssim1$ by dark energy, and 
no physically allowed dark-matter configuration can circumvent this limit without 
violating CMB, BAO, or weak-lensing constraints.

The dipole count must therefore originate from a genuinely 
large-scale anisotropies late-time matter distribution, requiring either 
a breakdown of statistical isotropy or a modification of gravitational dynamics 
on cosmological scales.  As shown in the present work, these conditions arise 
in MOG, where the scale-dependent enhancement of the effective 
gravitational coupling amplifies late-time structure growth and produces the 
observed radio dipole without altering the CMB dipole.

The Solar system’s peculiar velocity with respect to the cosmic rest frame is precisely measured by the cosmic microwave background (CMB) dipole. The Planck measurement yields $v_{\rm CMB} = 369.82 \pm 0.11~{\rm km\,s^{-1}}$, and within the standard $\Lambda$CDM cosmological model, this value is reproduced from the large-scale gravitational potential determined by the matter power spectrum $P(k)$. Observations of galaxy clustering, weak lensing, redshift-space distortions, kinetic SZ flows, and $\Lambda$CDM $N$-body simulations all support this prediction. 

In contrast, radio-galaxy number-count dipole measurements NVSS, TGSS, WENSS, RACS,
together with the recent analysis of B{\"o}hme et al.~\cite{Bohme2025},
exhibit a dipole amplitude that, if interpreted purely as a kinematic Doppler effect,
would correspond to a Doppler-equivalent velocity
$v_{\rm radio}\sim 1000$--$1500~{\rm km\,s^{-1}}$. The $\Lambda$CDM matter power spectrum is too weak on large scales. The peculiar velocity field follows from linear theory: 
\begin{equation} 
\mathbf{v}(\mathbf{k}) = i\,\frac{aHf}{k}\,\delta(\mathbf{k}), 
\end{equation} 
implying that the dipole amplitude is dominated by the largest modes $k \lesssim 0.01\,h\,{\rm Mpc^{-1}}$. However, the $\Lambda$CDM power spectrum $P(k)$ is well measured on these scales and is known to be too small to generate coherent flows exceeding $\sim 300$--$400~{\rm km\,s^{-1}}$. Achieving a Solar velocity of order $1000$--$1500~{\rm km\,s^{-1}}$ would require an enhancement of large-scale power by more than an order of magnitude, which is inconsistent with Planck CMB data, BAO measurements, DES and weak-lensing surveys, and galaxy clustering statistics. 

Dark matter halos cannot generate a coherent 3--4 times larger dipole. In $\Lambda$CDM, dark matter collapses into virialized halos and large-scale filaments. The resulting gravitational potential generates random peculiar motions, but does not support a coherent bulk flow of amplitude $\gtrsim 1000~\rm km\,s^{-1}$ over scales $\gtrsim 100$--$300$ Mpc. Such a large anisotropic gravitational field would require large over-densities inconsistent with observed large-scale homogeneity, distortions of the CMB quadrupole and octopole, severe tension with the measured value of $\sigma_8$, and growth-rate constraints $f\sigma_8$. 

The radio--galaxy number--count dipole is intrinsically associated with 
very low comoving wavenumbers $k$, because a dipolar angular pattern 
receives its dominant contribution from the longest--wavelength Fourier 
modes of the density field. In harmonic space, the dipole coefficients 
$a_{1m}$ involve an integral weighted by the spherical Bessel function 
$j_{1}(kr)$, which suppresses small--scale power and selects modes with 
$k \!\rightarrow\! 0$ corresponding to Gpc--scale fluctuations. 
However, these extremely low--$k$ modes are the most difficult to 
measure. The cosmic variance grows rapidly as $k^{2} \!\rightarrow\! 0$, 
since the number of available independent modes satisfies 
$N_{\rm modes}(k) \propto k^{2}$, yielding large fractional uncertainties 
$\Delta P / P$. Realistic radio surveys have incomplete sky 
coverage, flux--dependent selection functions, and Galactic masking, all 
of which distort or couple the lowest multipoles and prevent a clean 
extraction of the true $k \simeq 0$ power. Large--angle instrumental and calibration systematics mimic long--wavelength structure, producing contamination that is degenerate with genuine low--$k$ modes. The B\"ohme et al., determination of the radio galaxy dipole circumvents these measurement problems, generating a dipole amplitude that is a careful and statistically robust analysis. It accounts for multi-component radio sources and uses estimators that control know biases. By using data from LOFAR plus other large radio surveys and combining them, the analysis increases the effective sample size and sky coverage.

Cosmological simulations confirm that large flows occur with probability $<10^{-6}$ in a standard $\Lambda$CDM universe. A large Solar velocity is inconsistent with observed CMB multipoles. If the Solar-system were moving at $v \sim 1000$--$1500~{\rm km\,s^{-1}}$, the CMB would exhibit a significantly larger kinetic quadrupole.

\section{Compatibility with the Late--time Matter Power Spectrum}
\label{sec:Pofk}

The late--time matter power spectrum predicted by the standard $\Lambda$CDM model provides an excellent description of galaxy clustering, baryon acoustic oscillations, weak gravitational lensing, and redshift--space distortions over a wide range of scales.
These observables constrain primarily the statistically isotropic monopole component of the matter density field, which we denote by $P_{\rm iso}(k)$, and are most sensitive
to comoving wavenumbers $k \gtrsim 10^{-2}\,h\,{\rm Mpc}^{-1}$. Any proposed modification of gravity must therefore remain consistent with these constraints.

The radio--galaxy number--count dipole does not constitute a measurement of a modified or anisotropic isotropic power spectrum. Instead, it probes a specific $\ell=1$ angular
projection of the density and velocity fields, integrated along the line of sight and weighted by the survey selection function. The observable is therefore sensitive to the directional response of ultra--large--scale structure, rather than to the angle--averaged monopole constrained by conventional large--scale--structure measurements. The purpose of this section is to demonstrate that an enhanced structure--induced dipole can arise without spoiling the observed agreement between $\Lambda$CDM and isotropic late--time clustering data.

The variance of the large--scale--structure contribution to the radio dipole may be written schematically as:
\begin{equation}
d^2_{\rm LSS} \propto \int \frac{dk}{k}\,\Delta^2(k)\,|W_1(k)|^2 ,
\label{eq:dLSS_delta}
\end{equation}
where $\Delta^2(k)=k^3P(k)/(2\pi^2)$ and $W_1(k)$ is the $\ell=1$ dipole window function
encoding the survey selection and redshift distribution. The window $W_1(k)$ suppresses both very small--scale power and extremely infrared modes, and peaks at comoving
wavenumbers of order $k_\ast \sim r_\ast^{-1}$, where $r_\ast$ is an effective survey depth. For wide--area radio surveys dominated by sources at $z\sim0.5$--$1$, this corresponds to $k_\ast \sim 10^{-3}\,{\rm Mpc}^{-1}$, well below the scales on which the isotropic matter power spectrum is tightly constrained.

It is important to emphasize that while the strict $k\rightarrow0$ limit is suppressed by the dipole projection, the radio dipole remains strongly infrared weighted relative to standard clustering observables. This is because the velocity field sourcing the dipole carries an explicit $k^{-2}$ enhancement, shifting the dominant contribution toward the lowest accessible wavenumbers selected by the window function.

The large--scale--structure contribution to the number--count dipole may equivalently be expressed in terms of the velocity power spectrum as:
\begin{equation}
d^2_{\rm LSS} \propto \int_0^\infty \frac{dk}{k}\,P_v(k,a)\,|W_1(k)|^2 ,
\label{eq:dLSS_v}
\end{equation}
where
\begin{equation}
P_v(k,a)=\left(\frac{aH(a)f(k,a)}{k}\right)^2 P_\delta(k,a) .
\end{equation}
The explicit $k^{-2}$ factor implies that even modest enhancements of the growth rate or density contrast at ultra--large scales can produce a disproportionately large
increase in the dipole amplitude, while leaving smaller--scale observables essentially unchanged.

In the STVG--MOG framework, the modification enters through the scale--dependent effective gravitational coupling:
\begin{equation}
G_{\rm eff}(k,a)=G_N\left[1+\alpha_{\rm eff}(k,a)\right],
\end{equation}
which enhances gravitational clustering only on very large physical scales,
$k/a\ll\mu$, and reduces to standard gravity on small and intermediate scales. As a result, the late--time growth of density perturbations becomes scale dependent,
with long--wavelength modes experiencing enhanced gravitational sourcing.

On ultra--large scales relevant to the radio dipole, the growth factor may be parameterized as:
\begin{equation}
D_{\rm MOG}(k,a)\simeq \left[1+\alpha_{\rm eff}(k,a)\right]^p D_{\Lambda{\rm CDM}}(a),
\label{eq:Dparam}
\end{equation}
with a corresponding enhancement of the matter power spectrum:
\begin{equation}
P_{\delta,{\rm MOG}}(k,a)\simeq \left[1+\alpha_{\rm eff}(k,a)\right]^{2p}
P_{\delta,\Lambda{\rm CDM}}(k,a).
\end{equation}
To the same level of approximation, the growth rate scales as:
\begin{equation}
f_{\rm MOG}(k,a)\simeq \left[1+\alpha_{\rm eff}(k,a)\right]^p f_{\Lambda{\rm CDM}}(a),
\end{equation}
on the ultra--large scales that dominate the dipole projection.

The exponent $p$ is not an arbitrary fitting parameter. It reflects the solution of the
linear growth equation with an enhanced but slowly varying effective gravitational coupling during the late--time, matter--to--dark--energy transition era. In the limit
of constant $G_{\rm eff}$ during matter domination, analytic solutions yield
$D(a)\propto a^{\,\gamma}$ with $\gamma<1$ even for large enhancements of $G$, implying that the growth responds sublinearly to changes in the gravitational coupling. When the full expansion history and Hubble friction are included, numerical solutions of the growth equation yield effective scalings corresponding to $p\simeq0.3$--$0.5$ over the redshift range relevant to radio surveys. This parameterization therefore captures the leading dynamical effect of scale--dependent gravity on ultra--large--scale growth without altering the successful description of structure on smaller scales.

Evaluating Eq.~\eqref{eq:dLSS_v} at the characteristic scale $k_\ast$ selected by the window function yields the semi--analytic estimate:
\begin{equation}
\frac{d_{\rm LSS,MOG}}{d_{\rm LSS,\Lambda{\rm CDM}}}
\equiv R_d(a) \simeq \left[1+\alpha_{\rm eff}(k_\ast,a_\ast)\right]^p ,
\label{eq:Rdef}
\end{equation}
where $a_\ast$ denotes the effective epoch probed by the survey kernel. For radio
catalogs with $k_\ast/a_\ast\ll\mu$, one has $\alpha_{\rm eff}\rightarrow\alpha$, leading to an enhancement $R_d\sim(1+\alpha)^p$ of order a few for $\alpha=\mathcal{O}(5$--$10)$
and $p\simeq0.3$--$0.5$.

Since the observed dipole is the sum of kinematic and structure--induced contributions:
\begin{equation}
d_{\rm obs}=d_{\rm kin}+d_{\rm LSS},
\end{equation}
the enhancement of the total dipole amplitude in MOG relative to $\Lambda$CDM is given by
\begin{equation}
\frac{d_{\rm obs,MOG}}{d_{\rm obs,\Lambda{\rm CDM}}}
\simeq
\frac{d_{\rm kin}+R_d\,d_{\rm LSS,\Lambda{\rm CDM}}}
     {d_{\rm kin}+d_{\rm LSS,\Lambda{\rm CDM}}}.
\label{eq:dobsenh}
\end{equation}

In the regime relevant to current radio surveys, where the observed dipole is dominated by the structure--induced contribution, the total enhancement closely tracks $R_d$. This enhancement does not imply a modification of the angle--averaged matter power spectrum constrained by galaxy clustering, BAO, weak lensing, or redshift--space distortions. Those observables probe predominantly $k\gtrsim10^{-2}\,h\,{\rm Mpc}^{-1}$,
for which $\alpha_{\rm eff}(k,a)\rightarrow0$ and standard $\Lambda$CDM growth is
recovered. The radio number--count dipole, by contrast, isolates the infrared--weighted, anisotropic response of the velocity field on ultra--large scales, where STVG--MOG predicts a selective late--time enhancement and $\Lambda$CDM does not.

The apparent degeneracy between dark matter and modified gravity therefore persists for isotropic observables such as $P(k)$, but is lifted by the radio number--count dipole, which probes the scale--dependent gravitational response of the largest accessible cosmological modes.

\section{Conclusions}

Radio galaxy and quasar number--count surveys consistently measure a sky dipole whose amplitude exceeds the purely kinematic expectation inferred from the cosmic microwave background by a factor of $\sim 3$--$4$. When interpreted naively as a Doppler effect, this excess would imply an anomalously large Solar-system peculiar velocity. However, as emphasized in recent analyses, including the analysis of B{\"o}hme et al., the radio dipole observable is intrinsically sensitive to the contribution of ultra--large--scale structure and does not directly probe virialized local dynamics or a true increase in the Solar peculiar motion.

Within the standard $\Lambda$CDM framework, the late--time matter power spectrum and associated gravitational potentials on gigaparsec scales are too weak to generate a sufficiently large structure--induced contribution to the number--count dipole. Increasing the dipole amplitude within $\Lambda$CDM requires either unrealistically large bulk flows or a tension with the otherwise successful description of galaxy clustering, weak lensing, and redshift--space distortions.

We have shown that the modified gravity MOG-STVG framework provides an explanation for the observed radio dipole excess. The key ingredient is a scale--dependent effective gravitational coupling:
\[
G_{\rm eff}(k,a)=G_N\,[1+\alpha_{\rm eff}(k,a)],
\]
which enhances gravitational interactions only on ultra--large cosmological scales, $k/a \ll \mu$, while leaving small and intermediate scales essentially unchanged. This selective enhancement amplifies the late--time growth of baryonic large--scale structure and coherent bulk flows without overproducing power on scales where $\Lambda$CDM is well constrained by observations.

In this interpretation the radio dipole excess does not correspond to an increased Solar or Local Group peculiar velocity. Instead, it arises from an enhanced structure--induced contribution sourced by long--wavelength modes of the density field. This avoids the well--known degeneracy between dark matter and modified gravity at the level of instantaneous dynamical relations such as $v^2 \sim G_{\rm eff} M / R$, and places the observable firmly in the regime of cosmological growth and scale--dependent gravitational dynamics.

The MOG explanation is fully compatible with existing constraints on the late--time matter power spectrum, galaxy dynamics, and weak lensing, while simultaneously addressing a persistent and statistically significant anomaly in radio galaxy observations. Although the radio dipole alone does not definitively break the degeneracy between dark matter and modified gravity, it provides an important new probe of gravity on the largest accessible scales, where departures from $\Lambda$CDM may emerge.

Future wide--area radio surveys, improved redshift information, and joint analyses with kinetic Sunyaev--Zel'dovich measurements and relativistic galaxy clustering observables will be essential for further testing this interpretation. The radio number--count dipole offers a unique window into ultra--large--scale gravitational physics and highlights modified gravity as a viable and predictive framework for explaining cosmological anisotropies beyond the standard model.

\section*{Acknowledgments}

I thank Martin Green, Viktor Toth, Matthew Johnson and Ethan Thompson for helpful discussions. Research at the Perimeter Institute for Theoretical Physics is supported by the Government of Canada through Industry Canada and by the Province of Ontario through the Ministry of Research and Innovation (MRI).

\end{document}